\def\year{2021}\relax
\documentclass[letterpaper]{article} 
\usepackage{aaai20}  
\usepackage{times}  
\usepackage{helvet} 
\usepackage{courier}  
\usepackage[hyphens]{url}  
\usepackage{graphicx} 
\usepackage{graphicx}  
\usepackage{algorithm}
\usepackage{algpseudocode}
\usepackage{subfigure}
\usepackage{amsmath}
\usepackage{booktabs}
\usepackage{amssymb}
\usepackage{color}

\title{An Algebraic-Topological Approach to Processing Cross-Blockchain Transactions}
\author{Dongfang Zhao\\
dzhao@uw.edu 
}
 
\begin{document}

\maketitle

\begin{abstract}
The state-of-the-art techniques for processing cross-blockchain transactions take a simple centralized approach:
when the assets on blockchain $X$, say $X$-coins, are exchanged with the assets on blockchain $Y$---the $Y$-coins,
those $X$-coins need to be exchanged to a ``middle'' medium (such as Bitcoin) that is then exchanged to $Y$-coins.
If there are more than two parties involved in a single global transaction, 
the global transaction is split into multiple local two-party transactions, each of which follows the above central-exchange protocol.
Unfortunately, the atomicity of the global transaction is violated with the central-exchange approach:
those local two-party transactions, 
once committed, 
cannot be rolled back if the global transaction decides to abort.
In a more general sense, the graph-based model of (two-party) transactions can hardly be extended to an arbitrary number of parties in a cross-blockchain transaction.
In this paper, we introduce a higher-level abstraction of cross-blockchain transactions.
We adopt the \textit{abstract simplicial complex}, 
an extensively-studied mathematical object in algebraic topology,
to represent an arbitrary number of parties involved in the blockchain transactions.
Essentially, each party in the global transaction is modeled as a vertex and the global transaction among $n+1$ ($n \in \mathbb{Z}$, $n > 0$) parties compose a $n$-dimensional simplex.
While this higher-level abstraction seems plausibly trivial,
we will show how this simple extension leads to a new line of modeling methods and protocols for better processing cross-blockchain transactions.

\end{abstract}

\section{Introduction}

\subsection{Motivation}

Since its inception in the form of Bitcoin~\cite{bitcoin},
blockchain has drawn more attention increasingly,
both technical and social-economical.
The applicability of blockchains has been considerably expanded,
from the original digital currency to electronic medical records~\cite{hdai_trialchain}, to smart government~\cite{bc_government}, to scientific experiments~\cite{aalmamun_sc19}, and so on.
With the increasing number of fields adopting blockchains,
a natural question is raised:

\textit{How could we exchange data among different blockchains?}

This seemingly simple question, however, brings up a set of new technical challenges,
some of which have been studied for a long time in a slightly different form in fields like databases, distributed computing, and graph theory.
For instance, in databases we had encounter challenges of migrating data among different vendors;
fortunately, SQL and a few standards were devised for the migration.
As an analogue, we are in the pre-SQL era of blockchains.

To better understand the problem of cross-blockchain transactions,
we need to review some basic concepts in blockchains, databases, graph theory, and distributed computing,
all of which will be reviewed in the section of \textbf{Preliminaries}.
For now, we briefly review two state-of-the-art solutions to the problem of cross-blockchain transactions,
one from the database community and the other from the distributed computing community.
The former one breaks the global transaction into a sequence of two-party sub-transactions, and the latter one extends the two-phase-commit (2PC) protocol with a witness blockchain.
Both approaches have been criticized regarding their limitations:
the first one cannot guarantee the atomicity of the global transaction, and the second one can lead to a blocking status of the transaction when the coordinator node fails in the middle of the transaction.
We will discuss more on both methods in the section of \textbf{Case Study and Evaluations}.

\subsection{Proposed Approach}

We propose to model the cross-blockchain transactions with a \textit{simplicial complex}---a well-studied mathematical object in algebraic topology.
The main reason we chose a simplicial complex for the model is its ability to portrait the high-dimensional relationship among parties in the transaction.
Instead of dealing with only two vertices in a graph, a simplicial complex can comprise a (subset of) arbitrary-dimensional object, in our case, a transaction.
As a consequence, we will be able to design more efficient protocols for handling the transactions in practice and, more importantly,
to overcome the known limitations of existing approaches, as summarized in Table~\ref{tbl:comparison}.

Topological approaches are being increasingly applied to various aspects of blockchains, such as ransom data analysis~\cite{cakcora_ijcai20}.
To our knowledge, however, this is the first work on topological methods for the cross-blockchain transaction (CBT)---one of the most critical subsystems for the future widespread deployment of next-generation blockchains.


\section{Preliminaries}
\label{sec:review}

\subsection{Blockchains}
There are many definitions of blockchains.
What we present here is from the perspective of data structures and distributed systems.
A blockchain is a replicated \textit{linked lists},
each of which is usually deployed to a distinct machine.
Each element of the linked list is called a \textit{block};
the linked list on each machine has a starting element called the \textit{genesis block}---holding the initial data of the blockchain.
Each (non-genesis) block is linked to the previous block with a hash-value that is uniquely calculated by the metadata of the previous block---this is exactly why the blockchain claims to be tamper-evident: one cannot falsify the data without violating the hash-value (or, the so-called ``hash-lock'' in some articles).
The only allowed operation to the blockchain is \textit{append};
in theory, one cannot modify the existing data in any block.
Since the data on existing blocks are immutable,
the only way to update the data is to apply \textit{updates} in the consequent blocks.
Therefore, the non-genesis blocks, from the perspective of data management,
are just the holders of data updates.
Each block typically holds a batch of data updates,
from tens of records to hundreds of them, depending on the implementation of the blockchain.
The basic structure for these data updates is a \textit{transaction},
which we will review in the next subsection.

It is also possible for a blockchain, especially for public blockchains,
to have more than one decedent nodes that form \textit{forks} or \textit{branches} in the topology.
These forks will compete with each other to finally win the race for holding the longest valid blocks since the forking.
We will see how this property impacts the topological modeling in later sections.

\subsection{Transactions and Directed Acyclic Graphs}

The term transaction is somewhat overused in multiple fields,
and yet the one formulated decades ago in the database community is, arguably, still the most accepted one today.
A transaction is composed of at least one entity that carries out a series of operations (possibly only one) whose effect to the entities (or users) is either complete or none.
Put it in another way:
the operations in a transaction are \textit{all or nothing};
there is no partial complete.
More formally, it is called the \textit{atomicity} of a set of operations that a transaction must enforce.
In practice, many transactions are applied between two participants (also called parties, peers, users, depending on different fields).
Therefore, a graph with vertices and edges becomes a natural data structure to model (two-party) transactions,
where each vertex represents a participant, and the edge between two vertices indicates a \textit{relationship},
in this case, a transaction.
The most widely-used data structure is a graph with the edge with an arrow pointing at a vertex to indicate the \textit{flow} of the data change.
Usually, we also assume the vertices are in a  \textit{partially ordered set},
which means we do not allow a loop in such a graph.
A graph with the aforementioned edge direction and without a loop is called a \textit{directed acyclic graph} (DAG).

\subsection{Distributed Commit Protocols}

A DAG is sufficient to model a two-party transaction, as discussed above.
For a transaction involving at least three parties,
although we do not (yet) have a formal model,
there are various protocols to ensure the atomicity under certain conditions.
The most widely-used protocols include 2PC~\cite{2pc} and 3PC~\cite{3pc}, and recently EasyCommit~\cite{msadoghi_edbt18}.
These distributed commit protocols share the same spirit in splitting the cluster of parties into a \textit{coordinator} and \textit{participants}\footnote{A participant node can serve as a coordinator as well.} and then model the communication between two nodes just as an edge in the DAG,
which is unable to model a higher-degree of transactions where parties are more than three.
Therefore, what we need here is an extended graph whose edges can connect more than two vertices,
which is called a \textit{hypergraph}.
One important subset of a hypergraph is called \textit{abstract simplicial complex},
a well-studied mathematical object in algebraic topology.
A hypergraph becomes an abstract simplicial complex when all the subsets of an edge are considered as edges of the hypergraph.
Our focus in this paper will be abstract simplicial complex because of the nature of multi-party transactions:
any local transaction of the global transaction should also be atomic.

\subsection{Abstract Simplicial Complex}

\begin{figure}[!t]
    \centering
    \includegraphics[width=65mm]{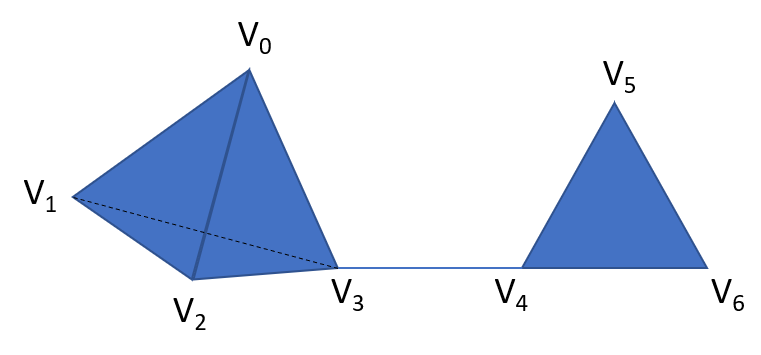}
    \caption{A simplicial complex $\sigma_3$ is composed of three simplices: a solid tetrahedron $\sigma_1 = \{v_0, v_1, v_2, v_3\}$, a fulfilled triangle $\sigma_2 = \{v_4, v_5, v_6\}$, and the newly introduced 1-simplex (i.e., an edge) $\{v_3, v_4\}$.
    }
    \label{fig:complex_example}
\end{figure}

We give a very brief review of abstract simplicial complex and related concepts in algebraic topology.
For full coverage of these, an introductory text is recommended, such as~\cite{jmunkres_book_at}.
We start with the concept of a \textit{simplex}.
A $n$-dimensional simplex $\sigma^n$ is a comprised of a set of $(n+1)$ elements (i.e., vertices),
where any subset of its $(n+1)$ vertices is defined as a \textit{face} of $\sigma_n$ and any face itself is again a (distinct) simplex.
Note that this definition is purely combinatorial: 
we are not interested in the concrete position of the vertices or the distance between the vertices.
Indeed, we can ``realize'' a combinatorial object with more concrete build-up,
which is called \textit{geometric realization}.
For example, a 3-dimensional simplex $\sigma_1$ with four vertices $\{v_0, v_1, v_2, v_3\}$ can be converted into a (solid) tetrahedron in $\mathbb{R}^3$ and a 2-dimensional simplex $\sigma_2$ with three vertices $\{v_4, v_5, v_6\}$ can be viewed as a (fulfilled) triangle in $\mathbb{R}^2$.
A \textit{simplicial complex} is a union of an arbitrary number of simplices as long as the intersection of simplices includes only the faces of the simplices,
and a \textit{abstract simplicial complex} emphasizes the combinatorial nature of the object.
For the objects we are interested in cross-blockchain transactions,
all the objects are combinatorial;
therefore, we in the following will use the term simplicial complex to indicate abstract simplicial complex.
For example, we can link\footnote{There is a special meaning of ``link'' in algebraic topology; we use link to simply indicate the connection between two simplices.} the two simplices with a new 1-dimensional simplex (i.e., an edge), as shown in Fig.~\ref{fig:complex_example}:
\[
\sigma_3 = \{\{v_0, v_1, v_2, v_3\}, \{v_4, v_5, v_6\}, \{v_3, v_4\} \}.
\]

In this case, $\sigma_3$ has three simplices: $\sigma_1$, $\sigma_2$, and the unnamed 1-dimensional simplex $\{v_3, v_4\}$.
The dimensionality, denoted by $Dim(\cdot)$ of a simplicial complex is defined as the highest dimension among all of its simplices;
therefore, we have $Dim(\sigma_3) = Dim(\sigma_1) = 3$ in this example.

An important related concept to the simplicial complex is \textit{Betti numbers},
usually denoted $\beta$'s,
which represents a series of integers indicating the ``holes'' at specific dimensions. 
Betti numbers are one of the most widely used \textit{topological invariants}---an important property when studying topological objects because these invariants do not change after continuous transformation.
Roughly speaking, a ``hole'' is a loop with the encompassed area empty.
For instance, in our example of Fig.~\ref{fig:complex_example},
the 0-dimensional Betti number $\beta_0 = 1$, meaning that we have one connected component and any $\beta_k = 0$, $k \in \mathbb{Z}$ and $k \geq 1$, because the simplicial complex apparently does not exhibit any holes at one or higher dimensions.

\section{Algebraic Topological Modeling}

\subsection{The Basics: Single Blockchains without Forks}

We start with some notations used modeling the blockchains.
We use $C$ to indicate a cluster of blockchains,
where $c_i$ indicates the $i$th blockchain.
Each blockchain comprises a series of blocks,
which are denoted by $b_i^j$'s---the $j$th block of the $i$th blockchain.
In system implementation, one also needs to index (a cluster of) machines, e.g., $m_k$ indicating the $k$th machine, either a virtual machine/docker or a physical server/workstation.
It should be clear that there is no universal relation between blockchains and machines:
for example, two distinct blockchains can be deployed to the same set of machines; they can also be deployed to two exclusively distinct sets of machines, etc.
To put it in another way, at the transaction processing level we are only interested in the blocks being involved from multiple blockchains;
whether the blocks (either from the same or different machines) are within the scope of the system implementation, and thus not our concern when modeling the transactions and their protocols.

\begin{figure}[!t]
    \centering
    \includegraphics[width=85mm]{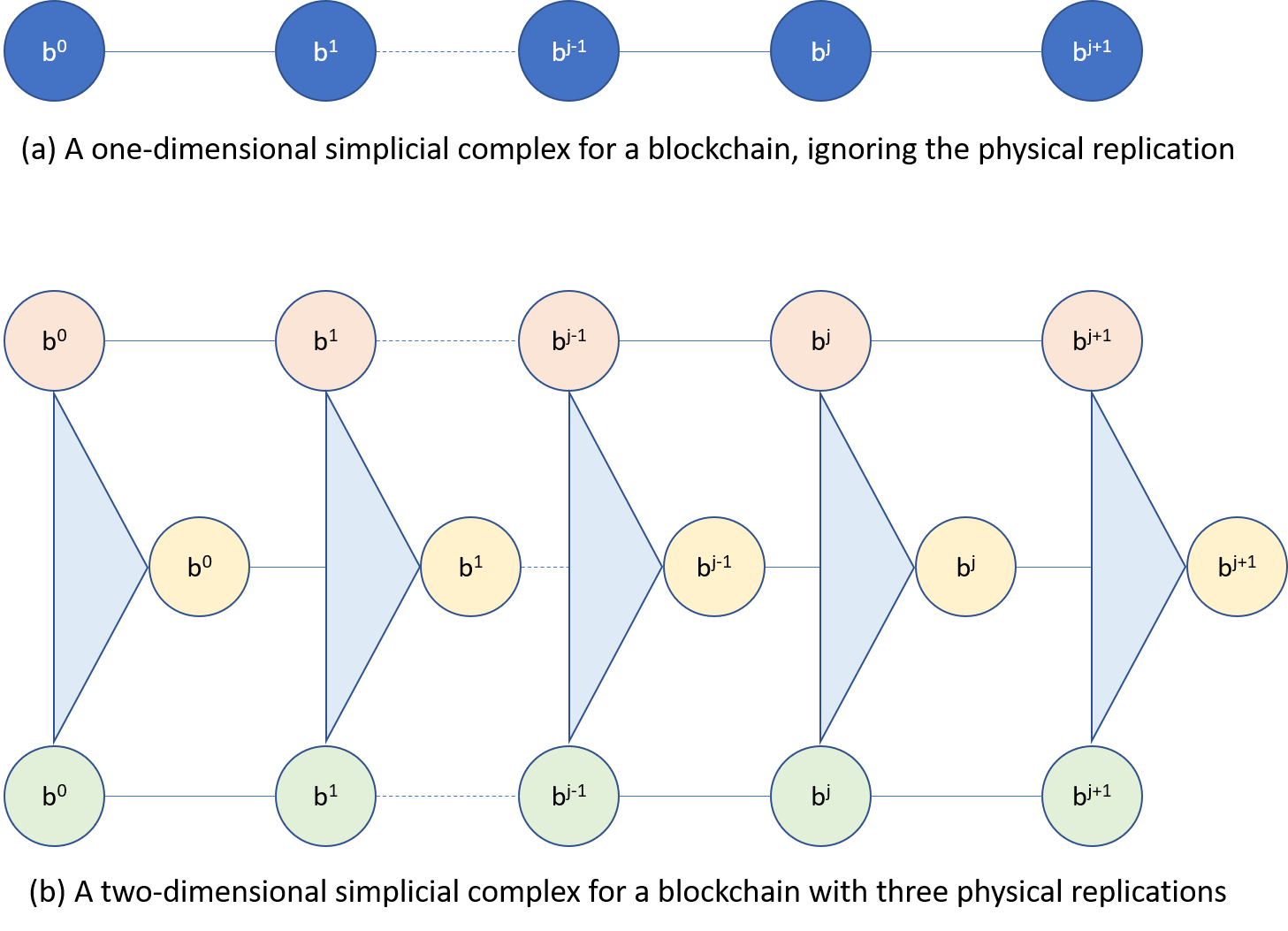}
    \caption{A blockchain can be abstracted into a one- or $k$-dimensional simplicial complex depending on whether to include the physical replication.
    }
    \label{fig:blockchain_complex}
\end{figure}

For the sake of clarity, in the following discussion, we simply abstract each blockchain as a single 1-dimensional simplicial complex,
just as a linked list.
In a more realistic setting, readers should keep in mind that a blockchain is really composed of many copies of such simplicial complex;
it is thus also reasonable, if not desirable, to model a blockchain as a $k$-dimensional simplicial complex if there are $(k+1)$ participants (i.e., $k+1$ copies on $k+1$ machines).
We illustrate the difference in Fig.~\ref{fig:blockchain_complex}:
the top diagram shows a simple one-dimensional simplicial complex (i.e., a linked list) without the physical replication detail;
the bottom diagram reveals that the blockchain has three participants, each of which holds a copy of the chain.
Formally, we can either model a single blockchain as a one-dimensional simplicial complex or an $\mathbf{m}$-dimensional simplicial complex to reflect the physical implementation,
where $\mathbf{m} = |\{m_k\}| - 1$.
Although in the remainder we will follow the first approach,
changing the assumption into the second approach would not invalidate the correctness and applicability of the proposed models and protocols.

\subsection{More Reality: Single Blockchains with Forks}

It is straightforward to extend the idea of topological modeling of single blockchains to the case where multiple forks appear.
In practice, a fork usually involves two branches,
and we, in the following text, will make such an assumption.
The discussion below, however, can be extended to three or more branches with simple changes.
Although a fork can persist indefinitely in theory,
practical blockchain systems can always employ some specific mechanisms to invalidate all but one fork so that the consequent blocks are unique among all machines (i.e., all replicas).
For example, Bitcoin re-checks, hourly, the lengths of two forks, and considers the longer one (i.e., with more blocks validated and appended) as the only valid chain.

\begin{figure}[!t]
    \centering
    \includegraphics[width=85mm]{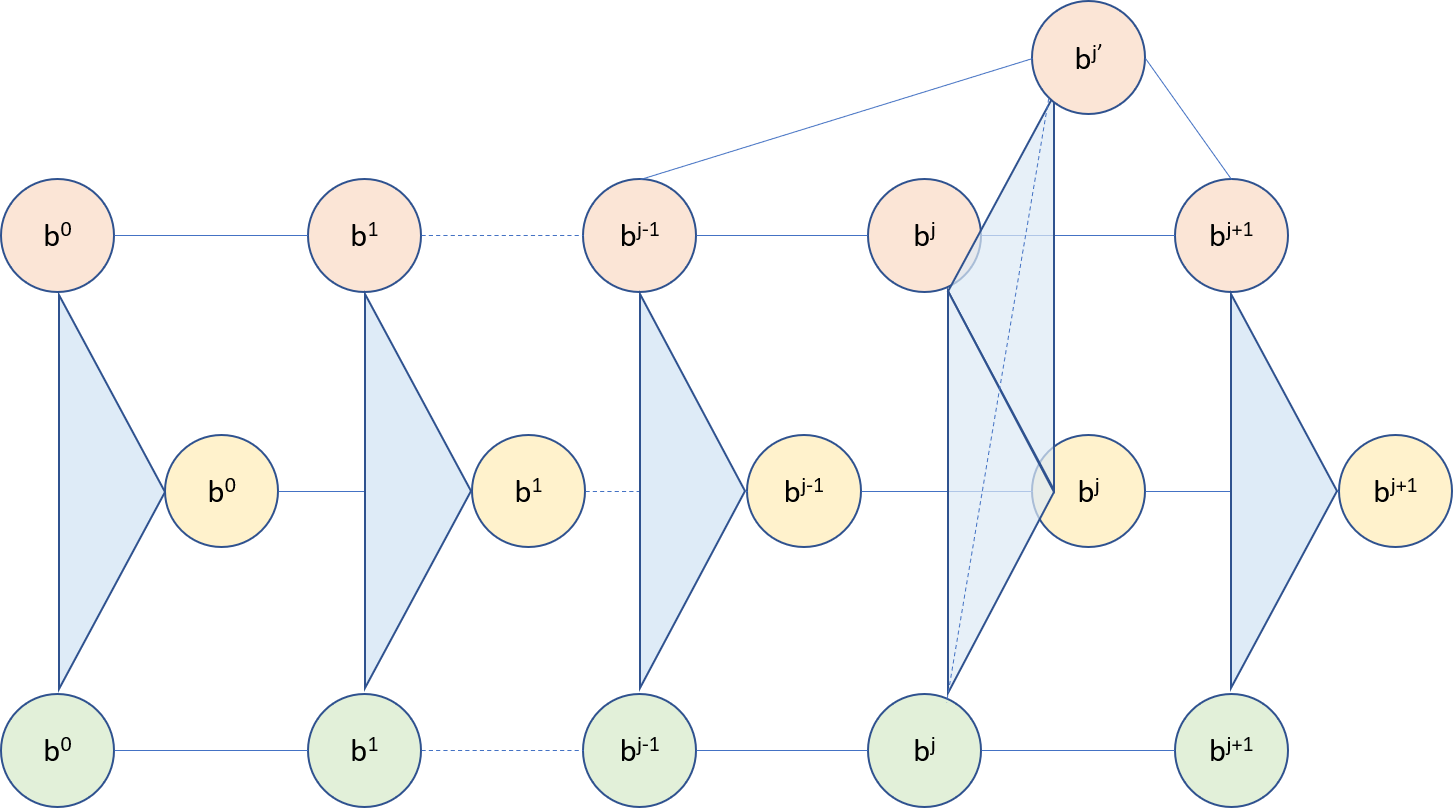}
    \caption{A single blockchain where one replication (on a specific machine) has one fork (block $b^{j'}$).
    As a result, the topology of this blockchain ends up with a three-dimensional simplicial complex.
    }
    \label{fig:single_blockchain_fork}
\end{figure}

To make matters more concrete, 
we illustrate how to extend the single-blockchain topology to forks in a real example that we have worked on before.
Fig.~\ref{fig:single_blockchain_fork} shows one of the simplest scenarios:
the first machine somehow spawns a fork at the $j$th block, denoted $b^{j'}$.
Fortunately, the blockchain quickly eliminates any ambiguity at the $(j+1)$th block.
Nonetheless, at the timestamp of block $j$,
we do observe four copies: two from the two forks of the first blockchain (pink color), the one each from the second (yellow color), and the third (green color) blockchain.

The only caveat here is the introduction of higher-dimensional simplices due to the fork.
We now have a three-dimensional simplex among three $b^j$'s and one $b^{j'}$, denoted as a blue, solid tetrahedron (the fourth blue object counting from the left). 
Indeed, we can still take the same approach we use for blockchains without forks to simplify this topology by \textit{shrinking} the replicas without a fork.
The reason we can only merge those replicas without forks is that the blocks are considered \textit{isomorphic},
which roughly says that we can map them back and forth with a continues map.\footnote{There is a more rigid mathematical definition of isomorphism; we refer the readers to a standard algebraic topology text for more detail.}

We have basically covered the cases for single blockchains,
now we are ready to discuss transactions among them.
As a convention, we will start with the ``civil'' cases where no forks appear.

\subsection{Perfect World: Multiple Blockchains without Forks}

If we assume that there is no fork at any involved blockchains,
we can simply model the clusters of machines where the blockchains are deployed as a single one-dimensional simplicial complex,
as exemplified by Fig.~\ref{fig:blockchain_complex}(a).
In this case, we can ignore the detailed replication of a single blockchain and put our focus on the inter-blockchain interaction on specific blocks.
As before, we start by introducing some notations.

An $n$-party transaction has $n$ distinct blockchains involved, indexed by $[1..n]$.
We assume that each of the $n$ blockchains has one and only one block containing the data involved in the $n$-party transaction;
these $n$ blocks are thus encoded as:
$\{b_1^{j_1}, \cdots, b_k^{j_k}, \cdots, b_n^{j_n}\}$,
where $j_k$ indicates the block index of the $k$th blockchain of the $n$-party transaction.
As a result, we can use an $(n-1)$-dimensional simplex among the $n$ blocks (i.e., vertices) $\{b_1^{j_1}, \cdots, b_k^{j_k}, \cdots, b_n^{j_n}\}$ to represent this $n$-party transaction.
When $n = 2$, the transaction is trivially modeled as an edge between two vertices, i.e., two blocks from two distinct blockchains.
Similarly, $n = 3$ implies an planar triangle, $n = 4$ implies an tetrahedron,
and $n \geq 5$ cannot be visually perceived. 
\begin{figure}[!t]
    \centering
    \includegraphics[width=85mm]{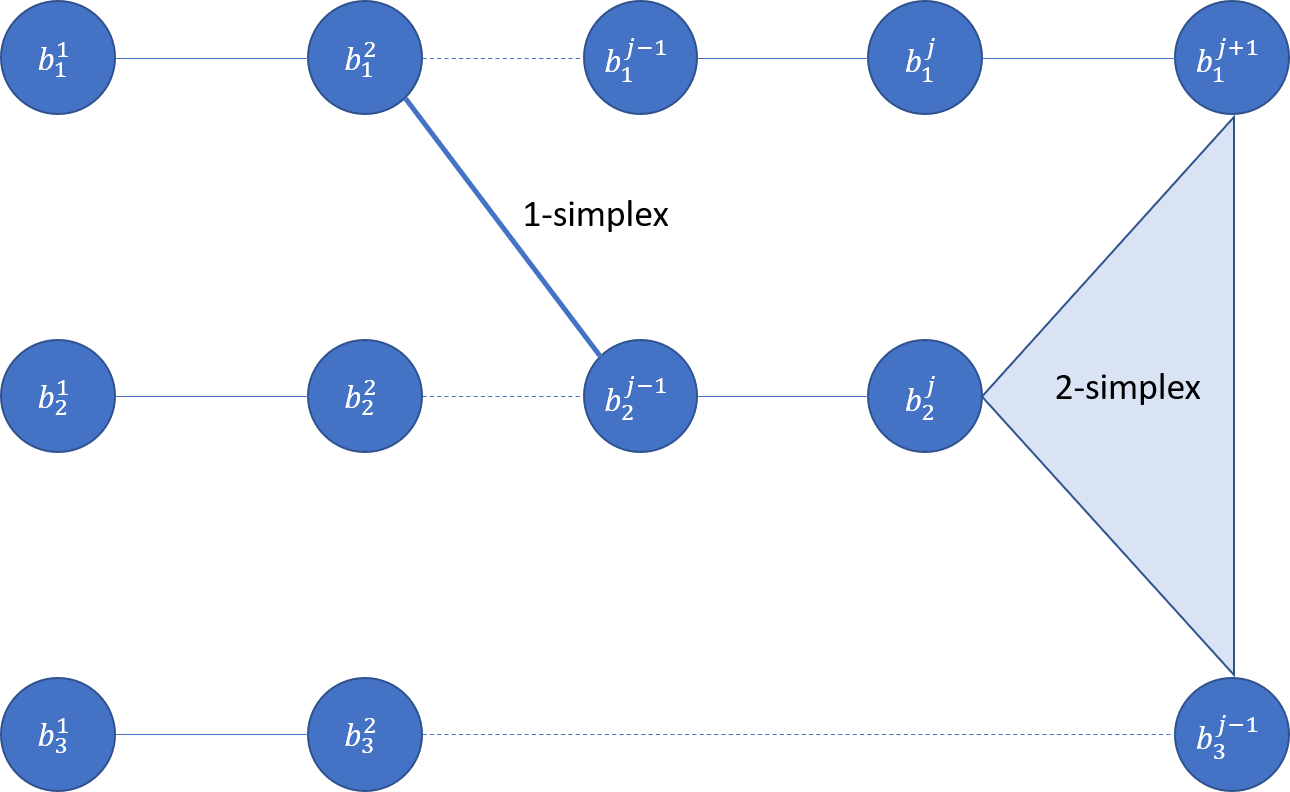}
    \caption{When no blockchain has any forks, transactions between two blockchains are 1-simplices,
    and transactions among three blockchains are 2-simplices.
    Note that the timelines (i.e., block appending rates) are not proportional to the lengths of chains.}
    \label{fig:txn_nofork}
\end{figure}

Fig.~\ref{fig:txn_nofork} shows a federation of three blockchains $c_1$, $c_2$, and $c_3$. 
For now, we are not interested in how many or what machines to which these three blockchains are deployed;
we will work on this matter in the next section.
There are two transactions $T_1$ and $T_2$ shown in the figure:
$T_1 (b_1^2, b_2^{j-1})$ and $T_2 (b_1^{j+1}, b_2^{j}, b_3^{j-1})$.

It should be noted that by the definition of simplex, 
any of its faces is also a simplex \textit{per se}.
Therefore, any of three edges (i.e., 1-simplices) in 2-simplex $T_2$ represents a relationship between the two endpoint blocks.
That is, $T_2$ implicitly encompasses a two-party transaction between $b_2^j$ and $b_3^{j-1}$ (and two more).

If we recall the naming convention of simplicial complexes algebraic topology, 
we can represent the topology of Fig.~\ref{fig:txn_nofork} as the following set:
\[
\{\{b_k^l, b_k^{l+1}\}, \{b_1^2, b_2^{j-1}\}, \{b_1^{j+1}, b_2^j, b_3^{j-1}\}\},
\]
where $k = [1,2,3]$ and $l \in \mathbb{Z}$, $l \geq 1$.
The first element represents the appending operation between adjacent blocks of a specific blockchain.
The second and third elements represent the two- and three-party transactions.

The topology in Fig.~\ref{fig:txn_nofork},
although only an oversimplification without forks,
exhibits interesting topological invariants.
For example, all of the three blockchains are correlated by some transactions---there is only one connected component.
Moreover, there is a cycle\footnote{There is a more formal definition of objects like this: it is called an \textit{oriented simplex} in algebraic topology.} (from the top-left corner, clockwise):
\[
[b_1^2, \dots, b_1^{j-1}, b_1^j, b_1^{j+1}, b_2^j, b_2^{j-1}],
\]
which is usually called a one-dimensional ``hole'' in algebraic topology.
Note that the loop $\{b_1^{j+1}, b_2^j, b_3^{j-1}\}$ is not a one-dimensional hole, because the the encompassed area is fulfilled.
Alternatively, the topology can be represented by its Betti numbers:
$\beta_0 = 1$ (one connected component), $\beta_1 = 1$ (one one-dimensional hole), and $\beta_k = 0, k \in \mathbb{Z}, k \geq 2$ (no two- or higher-dimensional holes.
As a side note, the Betti numbers are not trivially computable as in the example of Fig.~\ref{fig:txn_nofork};
in many, if not most, real-world applications, we need to design parallel algorithms to efficiently compute the matrices (in the vector spaces) of various algebraic groups,
which is beyond the scope of this paper.

\subsection{Wild West: Multiple Blockchains with Forks}

In practice, forks are very common, particularly in public blockchains, many of which are based on peer-competition, such as PoW~\cite{pow}.
We can extend our definitions and rules from previous sections to forks.
We will again start with a concrete example where only two blockchains are involved, and each has a fork (two branches) for the blocks of data in the transaction.

\begin{figure}[!t]
    \centering
    \includegraphics[width=85mm]{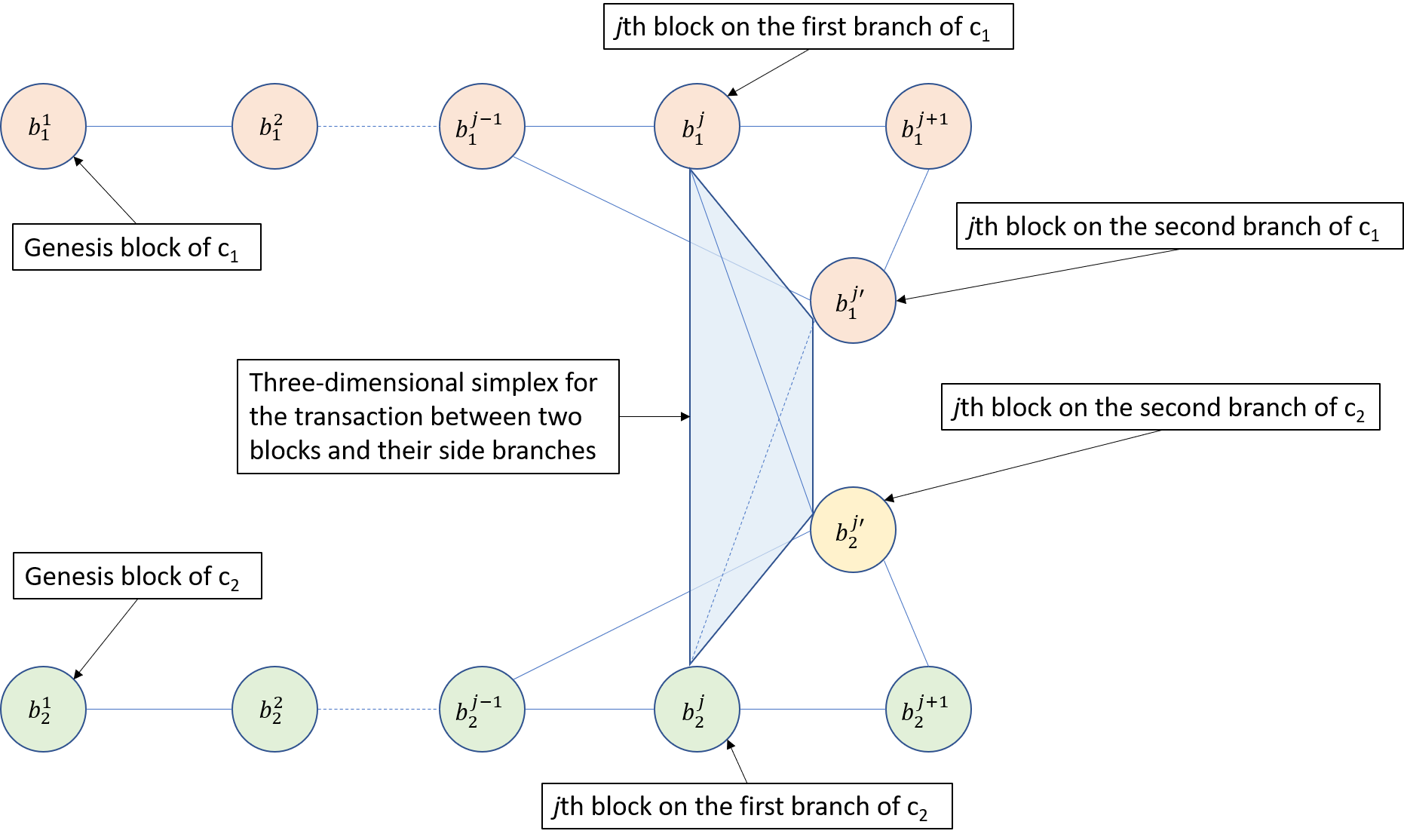}
    \caption{The topology of two distinct blockchains each of which exhibits two forks over the blocks involved in the two-party transaction.}
    \label{fig:txn_fork}
\end{figure}

Fig.~\ref{fig:txn_fork} illustrates that the transaction is between the $j$th block of blockchain $c_1$ and the $j$th block of blockchain $c_2$.
Because of the branches (i.e., forks) on both blocks,
the resulting simplex comprises four vertices,
thus a three-dimensional simplex.
Also, note that in this figure we did not show the possibly more replicated copies of the blocks (on distinct machines);
only two copies (branches) are shown for each blockchain.
The topological invariants of Fig.~\ref{fig:txn_fork} are:
$\beta_0 = 1$, $\beta_1 = 4$, and $\beta_k = 0$ for $k \geq 2 $ and $k \in \mathbb{Z}$.
To verify these, it is straightforward to see that all the blocks are connected,
and there are four one-dimensional holes:
$\{\{b_n^{j-1}, b_n^j, b_n^{j'}\},
\{b_n^j, b_n^{j+1}, b_n^{j'}\}\; |\; n = [1,2]
\}$.

In general, a $\mathbf{t}$-dimensional simplex is formed for the cross-blockchain transaction where $\mathbf{t}$ is calculated as: 
\[
\mathbf{t} = \sum_{i = 1}^\mathbf{\mathbf{n}} m_i + f_i,
\]
where $\mathbf{n}$ represents the total number of blockchains, $m_i$ represents the $i$th blockchain's total number of machines (or, replicas), and $f_i$ represents the $i$th blockchain's extra branches (forks) due to competition among machines.
We want to reemphasize that these Betti numbers will not be visually calculated, as in Fig.~\ref{fig:txn_fork} and might be very compute-intensive.
How to optimize the computation of Betti numbers is off the topic of this paper,
as we in this work focus on the topological modeling of cross-blockchain transactions and, more importantly, how to leverage this new modeling methodology to design more effective and efficient protocols to \textit{process} the transactions.
This is exactly what we will discuss in the next section.

\section{Topological Multi-Party Transaction}

Using the proposed topological model of the blocks from distinct blockchains,
we are able to design more effective protocols for committing and aborting transactions.
Before we present the protocol, we need a few more definitions to articulate the terms and flows.
For instance, we have repeatedly been talking about ``topology'',
which is also (over) used in other contexts,
but we never give a formal mathematical definition of what we mean by that.

\subsection{Terminology}

A \textit{topology} $\mathcal{T}$ of a set $S$ is a set of subsets of $S$, where $\emptyset$, $S$, and the union and finite intersection of any elements of $\mathcal{T}$ are also in $\mathcal{T}$.
For instance, if $S = \{a, b, c\}$,
then the set $\mathcal{T}_1 = \{\emptyset, S, \{a, b\}, \{b, c\}\}$ is not a topology of $S$ because 
\[
\{a, b\} \cap \{b, c\} = \{b\} \not\in \mathcal{T}_1.
\]
Obviously, the power set of $S$ is a topology of $S$ by definition (namely, the \textit{discrete topology}),
but it is not a very interesting construction:
it simply means all the elements can be arbitrarily joined together---no property or ``shape'' can be drawn from the data.

A closely related concept is \textit{topological space},
which is represented by the tuple $(\mathcal{T}_S, S)$;
literally, the topological space is a notation to indicate a set and its topology.
Indeed, a set can have many topologies, but a valid topology can have only one root set.
Therefore, people often use the terms topology and topological space interchangeably,
if no ambiguity arises.

Any element in a topology is called an \textit{open set},
a very common definition in any introductory algebraic topology text.
The definition is so commonly used that few texts really gives the intuitive idea behind its name:
what do we really mean by ``open''?
In fact, it has a more concrete meaning in real analysis (e.g., the definition of \textit{continuity});
in those definite sets which are the case of cross-blockchain transactions,
such ``openness'' implies the ``nearness'' relationship between blocks.
For example, 
if $S = \{a, b, c\}$,
and its topology $\mathcal{T}_S = \{\emptyset, S, \{a, b\}, \{b, c\}, \{b\}\}$,
then it topologically means $a$ is near $b$, $b$ is near $c$, but $a$ is not near $c$.
Any complementary set of an open set is called a \textit{closed set};
an element of the topology can be both an open and a closed set.
We will not discuss further on open and closed sets;
interested readers can refer to any introductory textbook on algebraic topology and real analysis.

\textit{Write-ahead logging} (WAL) is one of the most effective ways to ensure the consistency of data as well as to provide fault tolerance, 
which are closely related to the protocols for processing transactions. 
There are two major approaches to WAL: the \textit{UNDO} log and the \textit{REDO} log\footnote{Indeed, there is a hybrid approach mixing both UNDO and REDO; we do not discuss it since it is derived from the two basic approaches.}.
Essentially, a REDO log was appended when an early commit entry was persisted to the disk before the updates are applied;
the rationale is that if something bad happens to the updates,
the system can safely restart and re-run the operations from the point of REDO log.
On the other hand, an UNDO log is written \textit{before} the final commit entry was written to the disk;
if there is a failure before the commit,
the system will restart and roll back the partial changes.
The protocol proposed in this paper takes the second approach.

\subsection{Protocol}

We present the topological cross-blockchain transaction (TopoCBT) protocol in Protocol~\ref{alg:CBT}.
The input of the protocol includes the topology of the blockchains and the blocks involved in the transaction. 
In practice, the block can be retrieved by checking the public addresses of the sender and the receiver of the transaction and returning the index of the block that holds the transaction.
For instance, if the transaction involves three users Alice, Bob, and Cindy, 
represented by the notation $T(Alice, Bob, Cindy)$,
on three distinct blockchains $C_1$, $C_2$, and $C_3$,
then $B_T$ consists of the latest blocks (including the possible forks) that have the transactions touching Alice, Bob, and Cindy.

The output of Protocol~\ref{alg:CBT} is binary:
either all the requested operations pertaining $T$ are successfully completed (Line 15),
or none of them is persisted to the disk (Line 9).
If we recall the definition of \textit{atomicity} required by a transaction,
such a binary output is exactly what is desired. 
Indeed, there is more detail in the protocol,
as explained in the following.

\begin{algorithm}
\floatname{algorithm}{Protocol}
    \caption{Topological Cross-Blockchain Transaction}
    \label{alg:CBT}
    \begin{algorithmic}[1]
        \Require 
        A topology $\mathcal{T}$ representing the status of the blockchains.
        The blocks touched by the transaction are denoted by the set $B_T$, where $T$ indicates the transaction.
        \Ensure Either all the operations or none of them have been successfully completed; that is, atomicity is ensured.

        \Function{TopoCBT}{$\mathcal{T}$, $T$, $B_T$}
        
        \State Lock the blocks in $B_T$
        
        \State Construct the simplex $\sigma \in \mathcal{T}$ from blocks in set $B_T$
        
        \For {an open set $\delta \in \sigma$}
            \State Write the UNDO write-ahead log to the disk
            \State Try to apply updates to vertices in $\delta$
            \If {Updates unsuccessful} \Comment{Uncivil case}
                \State Write ``Abort'' to the disk
                \State Rollback the changes with the UNDO log
                \State Break down $\sigma$
                \State Release the lock on blocks of $B_T$
                \State \textbf{break} \Comment{Early termination}
            \EndIf
        \EndFor
        \State Write ``Commit'' to the disk \Comment{Civil case}
        \State Break down $\sigma$
        \State Release the lock on blocks of $B_T$          
        \EndFunction
    \end{algorithmic}
\end{algorithm}

To prevent the possible dirty reads from other threads,
the very first step of TopoCBT is to lock the related blocks $B_T$,
as shown in Line 2.
Then, based on the block indices from $B_T$,
we construct the simplex induced by $B_T$,
which, in this case, is a power set $2^{B_T}$ (Line 3).
Starting at Line 4, we enumerate all the open sets in the simplex for the transaction and write an UNDO log,
which can be used to roll back the partial change to the blocks (Line 9) if the updates to be applied are not eventually persisted (Line 7).
If failure did happen, the protocol would also clean up the status by breaking down the simplex and releasing the lock over it (Lines 10--12).
On the other hand, if everything runs smoothly (Lines 4--14), 
the protocol simply commits the transaction with the same clean-up procedure (Lines 16--17) as in the uncivil case.

As we can see,
except for the topological part whose semantic is somewhat foreign,
Protocol~\ref{alg:CBT} remains self-explanatory and straightforward for implementation:
the primitives such as locks, WAL logs, control loops, and condition checks all require only elementary programming experience. 
For example, there are no separate sub-protocols for a \textit{coordinator} and a set of \textit{participants} as in 2PC~\cite{2pc}.
We will have more to say and conduct a more comprehensive comparison between TopoCBT and other approaches in the following section.
Before the evaluation, we will provide some analysis of the proposed protocol.

Let $\mathbf{n}$ denote the cardinality of $B_T$, i.e., $|B_T| = \mathbf{n}$.
Line 2 takes $\mathcal{O}(\mathbf{n})$ and Line 3 takes $\mathcal{O}(\mathbf{n}^2)$.
Similarly, Line 16 takes $\mathcal{O}(\mathbf{n})$ and Line 17 takes $\mathcal{O}(\mathbf{n}^2)$.
The number of loops induced by Line 4 is determined by the number of sub-transactions within $T$, denoted $\mathbf{m}$.
Line 6 may take up to $\mathcal{O}(\mathbf{n})$.
Lines 10 and 11 take $\mathcal{O}(\mathbf{n})$ and $\mathcal{O}(\mathbf{n}^2)$, respectively, both for up to one time.
Therefore, the overall complexity is $\mathcal{O}(\mathbf{n}^2+ \mathbf{n}\mathbf{m})$.

\section{Case Study and Evaluation}

This section compares the proposed topological protocol with two other approaches to processing cross-blockchain transactions.
We will mostly take the famous car-trading example~\cite{mherlihy_podc18} for illustration purposes.
We first briefly review the car-trading problem.

The car-trading problem states that there are three users Alice, Bob, and Cindy,
who hold ETH (Ethereum), BTC (Bitcoin), and the (electronic) title of a Cadillac sedan, respectively.
Obviously, these three types of assets are managed respectively by three distinct blockchains: Ethereum, Bitcoin, and a blockchain designed for (the ownership of) car titles.
Someday, Alice decides to spend her ETH to buy Cindy's Cadillac;
but Cindy can only accept BTC,
so Alice asks Bob to exchange BTC with ETH and then complete the deal with Cindy.
This is exactly a three-party transaction across three blockchains:
one user wants to deal with two others simultaneously. 
Why can we not break it into three two-party transactions?
Because if Cindy walks away after the transaction between Alice and Bob is complete,
Alice would consider the transaction in an unexpected status:
it is not Alice's intention to invest in BTC, and even worse, the transaction is not atomic---only part of it (one-third) is done.
The question is: how can we ensure the atomicity of this three-party transaction?

In the original paper~\cite{mherlihy_podc18} where the car-trading problem was raised, the author proposed to break the multi-party transaction into a series of two-party sub-transactions, each of which is called an \textit{atomic cross-chain swap} (AC2S).
As the name implies,
AC2S is not really a transaction,
but only a replacement with local atomicity between the pair of adjacent parties.
Specifically, AC2S sets a time clock between every pair of parties in the local (sub)transaction;
if either party cannot meet the requirement specified by the sub-transaction in time\footnote{This is implemented by the \textit{smart contract}, which is not discussed in this paper.},
then the party who is late will be ``worse-off''---the asset she or he transferred earlier cannot be returned.
Therefore, not only the atomicity is violated in the sense of global transaction,
but also data can end up in an inconsistent status (i.e., financial loss). 
Nonetheless, the mechanism posed by AC2S is useful when time is not highly sensitive and the financial penalty is acceptable.

Later on in~\cite{vzakhary_vldb20}, AC3WN was proposed as a protocol supporting the full atomicity among blockchains.
AC3WN was extended from 2PC with the introduction to a \textit{witness blockchain} as the coordinator for managing the sub-transactions.
AC3WN claimed to deliver constant latency for multi-party transactions in terms of the number of messages,
although it was not implemented or evaluated with any real-world workload.
The overall complexity was not provided in the paper,
but following the $\mathbf{m}$ (number of sub-transactions) and $\mathbf{n}$ (number of blockchains and their block forks) notations,
the overall time complexity would be serialized to
$\mathcal{O}(\mathbf{n}^2 + \mathbf{m}^2)$,
where the first term represents the transaction processing among distinct blockchains and the second term implies the additional computational overhead incurred at the witness blockchain.
We only calculate and compare the serialized complexity because otherwise, it would not be fair comparison---Protocol~\ref{alg:CBT} was presented in a serial fashion but can be trivially parallelized. 
In fact, to the best of our knowledge, there was not yet any real blockchain system with cross-chain transactions fully implemented,
although a simulation work was recently reported in~\cite{xwang_blockchain20}.
Moreover, an additional spatial overhead of $\mathcal{O}(\mathbf{m})$ is expected due to the introduction of the witness blockchain.
One issue with AC3WN is that 2PC is a blocking protocol, and AC3WN did not overcome this limitation in its design.
Nonetheless, AC3WN remains the first distributed commit protocol for cross-blockchain transactions.

\begin{table}[t]
\small
    \caption{Comparison between TopoCBT and others.}
    \label{tbl:comparison}
    \centering
    \begin{tabular}{ lccc }
        \toprule
                    & AC2S          & AC3WN         & TopoCBT     \\ 
        \midrule
        Atomicity   & $\times$      & $\checkmark$  & $\checkmark$ \\ 
        Nonblocking & $\checkmark$  & $\times$      & $\checkmark$ \\
        Complexity   & $\mathcal{O}(\mathbf{m}\mathbf{n}^2)$  & $\mathcal{O}(\mathbf{n}^2 + \mathbf{m}^2)$      & $\mathcal{O}(\mathbf{n}^2 + \mathbf{m}\mathbf{n})$ \\
        Space Overhead    & $\mathcal{O}(1)$  & $\mathcal{O}(\mathbf{m})$      & $\mathcal{O}(1)$ \\
        Portability   & medium  & low      & high \\
        \bottomrule
    \end{tabular}
\end{table}

Table~\ref{tbl:comparison} compares some key features of the mainstream protocols for cross-blockchain transactions.
As we can see, the proposed TopoCBT is not the fastest protocol in the worst case due to the somewhat expansive construction of the algebraic topology,
and yet is superior in all of the remaining criteria.
For example, we list the portability of AC3WN as ``low'' due to its requirement of the additional witness blockchain,
which might be infeasible in many applications:
while the idea is neat and simple, a new servicing blockchain is far from trivial to implement.
In contrast, AC2S only requires smart contracts and time clocks,
or even better, TopoCBT can be applied to any blockchains as long as the commonly seen programming primitives are available such as locks and logs.
Even for the time complexity, TopoCBT could be comparable to AC3WN in practice,
despite the worst-case time performance.
The key insight is that the number of sub-transactions $\mathbf{m}$ is usually much smaller than the number of replicas of a blockchain.
To illustrate this, recall the car-trading problem,
where $m = 1$ but $n$ could be tens of thousands:
as of the writing of this paper, Bitcoin has 10,426 nodes worldwide~\cite{bitnodes}.
Therefore, the time complexity of both AC3WN and TopoCBT degenerates to $\mathcal{O}(\mathbf{n}^2)$, in a practical sense.

\section{Conclusion and Future Work}

Topological approaches are being increasingly applied to various aspects of blockchains, such as ransom data analysis~\cite{cakcora_ijcai20}.
This paper represents the first work on topological methods for the cross-blockchain transaction (CBT)---one of the most critical subsystems for the widespread deployment of next-generation blockchains from different domains.
We detail the proposed models and protocols for CBT backed by abstract simplicial complexes that have been extensively studied in algebraic topology.
Our analysis and case study reveals that the proposed protocol, namely TopoCBT,
achieves all the desired properties such as atomicity and nonblocking without sacrificing efficiency when comparing with the state-of-the-art.
It is our hope that this paper conveys a clear message that the introduction of an algebraic topological approach would fundamentally upgrade the line of tools we currently have for dealing with cross-blockchain transactions.

Due to the limited space, there are many more facets regarding the topological properties exhibited from cross-blockchain transactions that are not discussed in this paper.
Such properties include persistent homology, homotopy groups, cohomology rings, and many more.
This paper only employs the very basic concepts related to abstract simplicial complexes. 
In the future, we plan to continue exploring the potential of topological techniques applicable to blockchains and optimizing the protocols with a higher degree of parallelism.

\bibliographystyle{aaai}
\bibliography{ref_new}

\end{document}